\documentclass[journal]{IEEEtran}

\usepackage[pdftex]{graphicx}
\usepackage{amsmath,bm}
\usepackage{amssymb}

\begin{document}
\title{Importance of the window function choice for the predictive modelling of memristors}

\author{Valeriy~A.~Slipko and Yuriy~V.~Pershin,~\IEEEmembership{Senior~Member,~IEEE}
\thanks{V.~A.~Slipko is with the Institute of Physics, Opole University, Opole 45-052, Poland}
\thanks{Y.~V.~Pershin is with the Department of Physics and Astronomy, University of South Carolina, Columbia, SC 29208 USA (e-mail: \mbox{pershin@physics.sc.edu}).}
\thanks{Manuscript received November ..., 2018; revised ....}}

\maketitle

\begin{abstract}
Window functions are widely employed in memristor models to restrict the changes of the internal state variables to specified intervals.
Here we show that the actual choice of window function is of significant importance for the predictive modelling of memristors.
Using a recently formulated theory of memristor attractors, we demonstrate that whether stable fixed points exist depends on the type of window function used in the model.
Our main findings are formulated in terms of two memristor attractor
theorems, which apply to broad classes of  memristor models.
As an example of our findings, we predict the existence of stable fixed points in Biolek window function memristors and their absence in memristors described by the Joglekar window function, when such  memristors are driven by periodic alternating polarity pulses. It is anticipated that the results of this study will contribute toward the development of  more sophisticated models of memristive devices and systems.
\end{abstract}


\IEEEpeerreviewmaketitle

\section{Introduction} \label{sec:Intro}

During the past decade there have been many publications using window function-based models~\cite{strukov08a,Benderli09a,joglekar09a,Biolek2009-1,Prodromakis11a,Kvatinsky13a,ANUSUDHA2018130,yu2013memristor,takahashi2015spice,abdel2015memristor,Zha16a,georgiou16a} to describe the response of either discrete memristors or their circuits.
While in many cases such models provided reliable predictions, there are situations when the behavior of the circuit depends critically on the choice of window function.
It seems that there is little awareness of this fact as, typically, little attention is paid to picking the window function.
The purpose of this Letter is to show generally, and illustrate through specific examples, that the choice of window function is of significant importance for the predictive modelling of memristors.

To proceed, we shall first introduce memristive systems~\cite{chua76a} or, simply, memristors, and window functions.
Current-controlled memristive systems are defined by~\cite{chua76a}
\begin{eqnarray}
V_M(t)&=&R_M\left(\boldsymbol{x},I \right)I(t), \label{eq1}\\
\dot{\boldsymbol{x}}&=&\boldsymbol{f}\left(\boldsymbol{x},I\right), \label{eq2}
\end{eqnarray}
where $V_M$ and $I$ are the voltage across and current through the system, respectively, $R_M\left( \boldsymbol{x}, I\right)$ is the memristance (memory resistance), $\boldsymbol{x}$ is a vector of $n$ internal state variables, and $\boldsymbol{f}\left(\boldsymbol{x}, I \right)$ is a vector function~\footnote{Voltage-controlled memristive systems are defined similarly~\cite{chua76a}.}.
The window function $g_i\left( x_i,I\right)$ is a multiplicative factor (normally, $0 \leq g_i(x_i,I)\leq 1$) that  enters into the $i$th component of $\boldsymbol{f}\left(\boldsymbol{x}, I \right)$.
To ensure a zero drift of $x_i$ across the boundaries, the window functions take zero values at the boundaries, which are often located at $x_i=0$ and $x_i=1$.
While it is generally believed that the window functions describe certain physical processes in real memristors, the present authors are not aware of any derivations of window functions either from  fundamental physical theories or first-principles modelling.
Therefore, at the present level of knowledge, the window functions are rather representations of what we think about how memristors work, and these representations may be close or not so close to the reality.

\begin{figure}[b]
\centering \includegraphics[width=0.9\columnwidth]{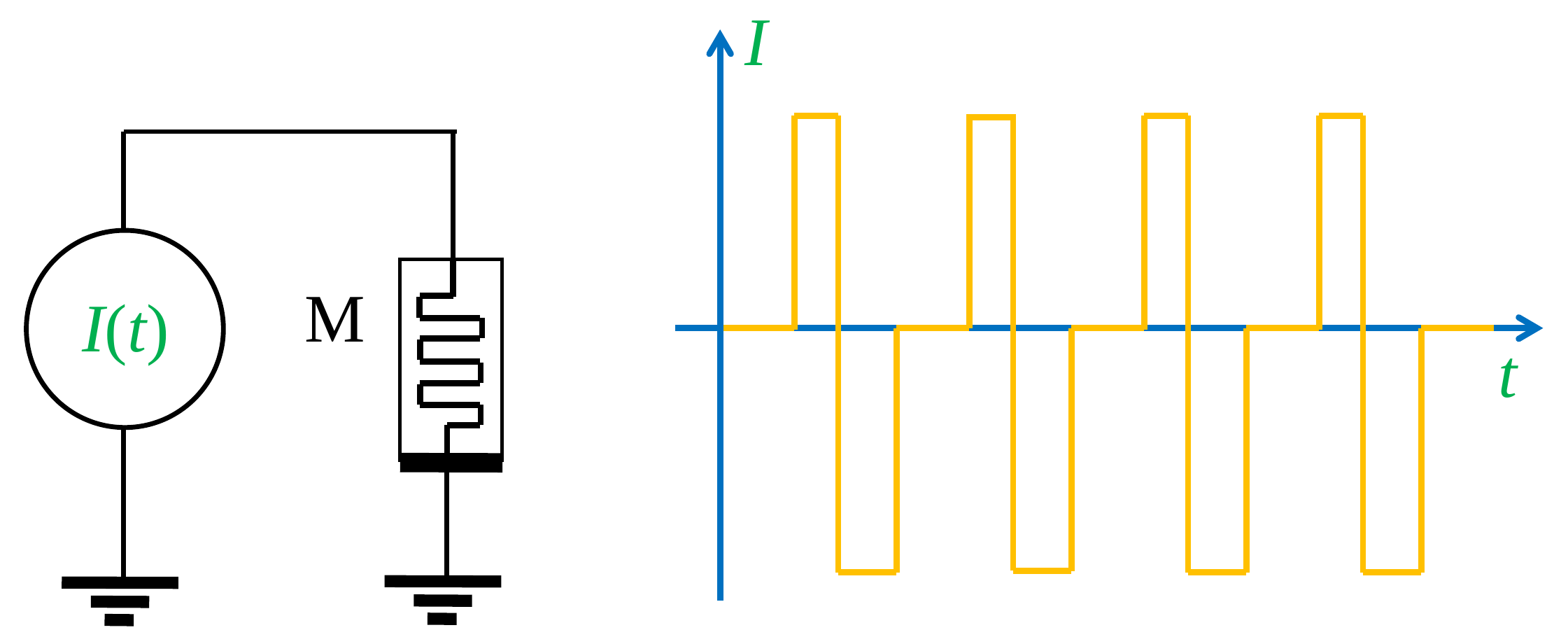} \\
(a)\;\;\;\;\;\;\;\;\;\;\;\;\;\;\;\;\;\;\;\;\;\;\;\;\;\;\;\;\;\;\;\;\;\;\;\;\;\;\;\;\;\;\;\;\; (b)
\caption{Circuit schematics (a) and pulse sequence diagram (b).
It is assumed that the memristor is subjected to narrow periodic pulses of alternating polarities, amplitudes $I_+$ and $I_-$, and durations $\tau_+$ and $\tau_-$, respectively.
}
\label{fig:1}
\end{figure}

In what follows we demonstrate that very close memristor models (differing only in their window functions) may exhibit qualitatively different dynamics. Specifically, we will analyze how the choice of window function influences the existence of stable fixed points (attractors) in driven memristors~\cite{pershin18b}.
Fig.~\ref{fig:1}(a) shows the circuit configuration considered in the present study.
Here, a single memristor is directly connected to a current source, which drives alternating polarity current pulses  through the memristor (the pulse sequence is sketched in Fig.~\ref{fig:1}(b)).
It is  shown that the presence or absence of stable fixed points in the time-averaged evolution of driven memristors can be directly related to the type of window function used in the memristor model.
For instance, we show that when using the Biolek window function, memristors do have stable fixed point dynamics, but not when using the Joglekar window function, even though all other conditions are identical.
We thus argue that the predictive modelling of memristors and their circuits requires a further refinement of memristor models in general and  window functions in particular.
It is important to note that the presence or absence of stable fixed points in driven memristors can be easily verified experimentally with physical memristors, and conclusions regarding the suitability of the use of a particular memristor model can then be made in particular cases.

The rest of this paper is organised as follows.
In Section~\ref{sec:2} (Preliminaries) we introduce the Joglekar and Biolek window functions, as well as stable fixed points in driven memristors.
The stable fixed points in two broad classes of memristor models are investigated in Section~\ref{sec:3}, which formulates our main findings in terms of two memristor attractor theorems.
In the same section, the theorems are exemplified by Biolek and Joglekar window function memristors.
Some concluding remarks are made in Section~\ref{sec:4}.

\section{Preliminaries} \label{sec:2}

\subsection{The window functions used in this paper}

This subsection briefly introduces the window functions used in the present paper.
As was mentioned above, window functions~\cite{strukov08a,Benderli09a,joglekar09a,Biolek2009-1,Prodromakis11a,Kvatinsky13a,ANUSUDHA2018130,yu2013memristor,takahashi2015spice,abdel2015memristor,Zha16a,georgiou16a}  are frequent components of memristor models~\cite{Kvatinsky13a,Kvatinsky15a}.
Typically, their role is to slow down the change of $x$ when the internal state variable $x$ approaches a boundary value (such as 0 or 1).

\begin{figure}[tb]
\centering \includegraphics[width=70mm]{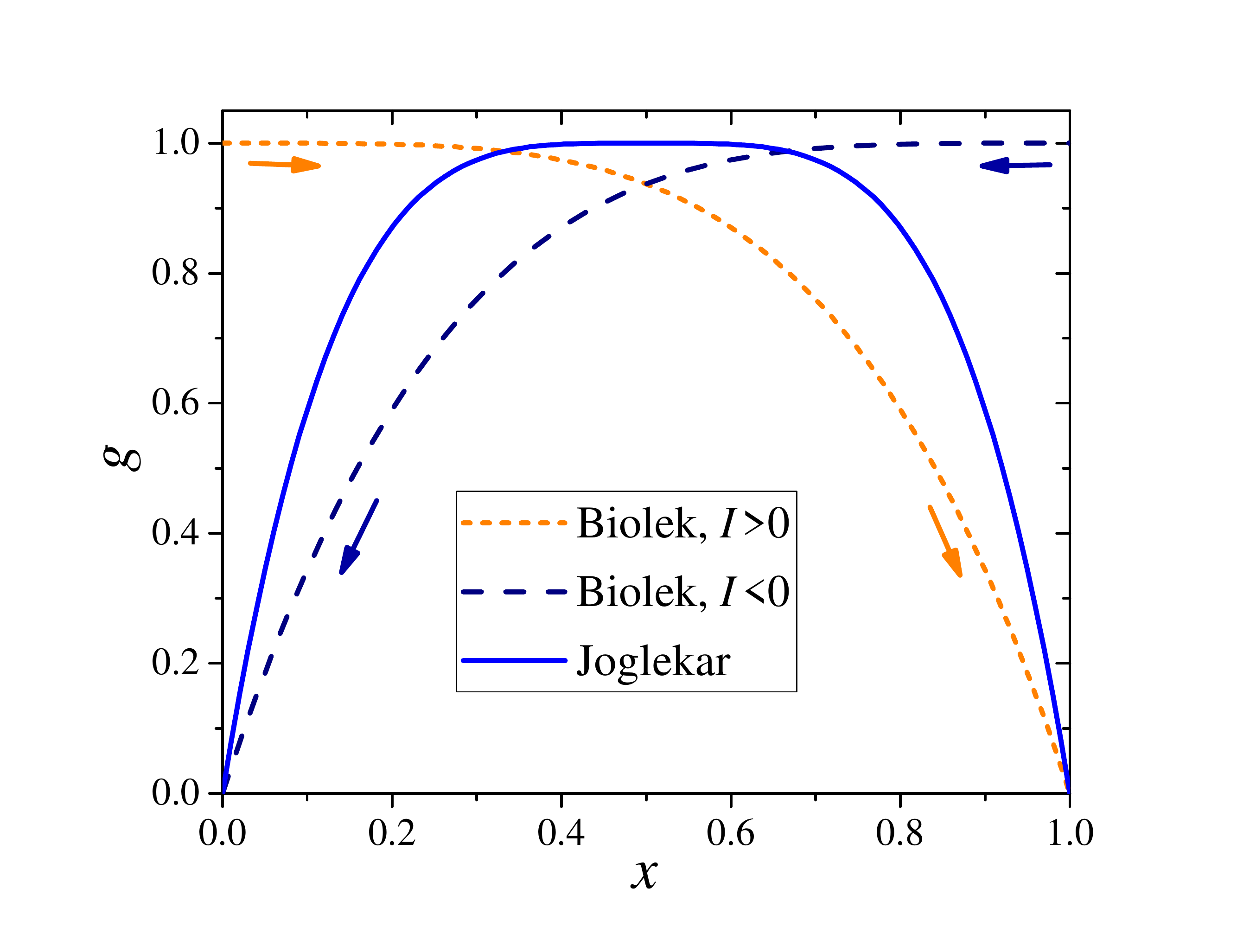}
\caption{Joglekar (Eq.~(\ref{eq:joglekar})) and Biolek (Eq.~(\ref{eq:biolek})) window functions plotted for $p=2$.
The arrows show the direction of  internal state dynamics in the Biolek model.}
\label{fig:2}
\end{figure}

In particular, the Joglekar window function $g_J(x)$ is defined by \cite{joglekar09a}
\begin{equation}\label{eq:joglekar}
g_J(x)=1-\left( 2x-1 \right)^{2p},
\end{equation}
where $p$ is a positive integer.
In the Biolek approach~\cite{Biolek2009-1}, the window function is given by
\begin{equation}\label{eq:biolek}
g_B(x,I)=1-(x-H(-I))^{2p},
\end{equation}
where $H(I)$ is the Heaviside step function and $p$ is a positive integer.
Fig.~\ref{fig:2} presents the Joglekar and Biolek window functions plotted for $p=2$.
Note that the Biolek window function (as a function of $x$) is different for positive and negative currents, and thus has a finite discontinuity at $I=0$.

\subsection{Stable fixed points in driven memristors}

Recently, the present authors  demonstrated the possibility of attractor dynamics in the time-averaged evolution of periodically-driven
memristors and memristive networks~\cite{pershin18b}. Here, we provide a detailed derivation of
relevant equations.
Consider a first-order current-controlled memristive system ($x$ is a scalar and $f(x,I)$ is a scalar function) connected to a current source and subjected to narrow alternating polarity current pulses (see Fig.~\ref{fig:1}). Let $\bar{x}$ denotes the internal state variable time-averaged over the pulse period $T$ according to
\begin{equation}\label{eq:average}
  \bar{x}(t)=\frac{1}{T}\int\limits_{t}^{t+T} x(\tau)\textnormal{d}\tau.
\end{equation}
The time derivative $\dot{\bar{x}}(t)$ evaluated using the Leibniz integral rule is given by
\begin{equation}\label{eq:average1}
  \dot{\bar{x}}(t)=\frac{x(t+T)-x(t)}{T}.
\end{equation}
Integrating Eq. (\ref{eq2}) from $t$ to $t+T$ under the assumption of a small change in $x$ by each pulse
(so that $x$ can be replaced by $\bar{x}$ in $f(x,I)$) we find
\begin{equation}\label{eq:average2}
 x(t+T)-x(t)= f(\bar{x},I_+)\tau_++f(\bar{x},I_-)\tau_-+f(\bar{x},0)\tau_0,
\end{equation}
where $\tau_0=T-\tau_+-\tau_+$. In what follows, we set $f(x,0)=0$ typically satisfied in the
first-order non-volatile memristor models.
Consequently, the equation for the time-averaged evolution takes the form
\begin{equation}\label{eq:average3}
  \dot{\bar{x}}(t)=\frac{1}{T}\left(f(\bar{x},I_+)\tau_++f(\bar{x},I_-)\tau_-\right).
\end{equation}

It follows from Eq. (\ref{eq:average3}) that the condition for the existence of a fixed-point attractor at $x_a$ can be written as~\cite{pershin18b}
\begin{equation}
f(x_a,I_+)\tau_++f(x_a,I_-)\tau_-=0   \label{eq:3}.
\end{equation}
Moreover, the fixed point is stable when~\cite{pershin18b}
\begin{equation}
\left.\frac{\partial f(x,I_+)}{\partial x}\right|_{x=x_a}\tau_++ \left.\frac{\partial f(x,I_-)}{\partial x}\right|_{x=x_a}\tau_-<0,  \label{eq:5}
\end{equation}
where $I_+>0$, $I_-<0$, $\tau_+$, and $\tau_-$ are the pulse parameters (defined in Fig.~\ref{fig:1}).

We also note that such periodically-driven memristors can be described by a
 memristor potential function~\cite{pershin18b}
\begin{equation}\label{eq:6}
U(x)=-\int \left[ f(x,I_+)\tau_++f(x,I_-)\tau_- \right] \textnormal{d}x,
\end{equation}
so that the problem of finding the attraction points (based on
 Eqs.~(\ref{eq:3}) and (\ref{eq:5})) corresponds to the problem of potential function
 minimisation.
To put it differently, every minimum of the potential function $U(x)$ represents a stable fixed point.

\section{Memristor dynamics}  \label{sec:3}

\subsection{Theorem 1} \label{sec:3a}

We first formulate and prove a theorem concerning the occurrence of stable fixed points in the memristors described by Eq.~(\ref{eq2}) of the form
\begin{equation}
  \dot{x}=h(I)g(x,I), \label{eq:bio_f}
\end{equation}
where \textit{i}) $g(x,I)$ is a continuous function of $x$;
\textit{ii})   $h(I)\geq 0$, $g(x,I)$ is monotonically decreasing as a function of $x$ with $g(0,I)=1$ and $g(1,I)=0$ for $I>0$; \textit{iii})  $h(I)\leq 0$, $g(x,I)$ is monotonically increasing as a function of $x$ with $g(0,I)=0$ and $g(1,I)=1$ for $I<0$.
We emphasise that $g(x,I)$ in Eq.~(\ref{eq:bio_f}) represents a quite general class of functions that includes but not limited to the Biolek window function given by Eq.~(\ref{eq:biolek}).

\begin{figure}[t]
\centering (a) \includegraphics[width=70mm]{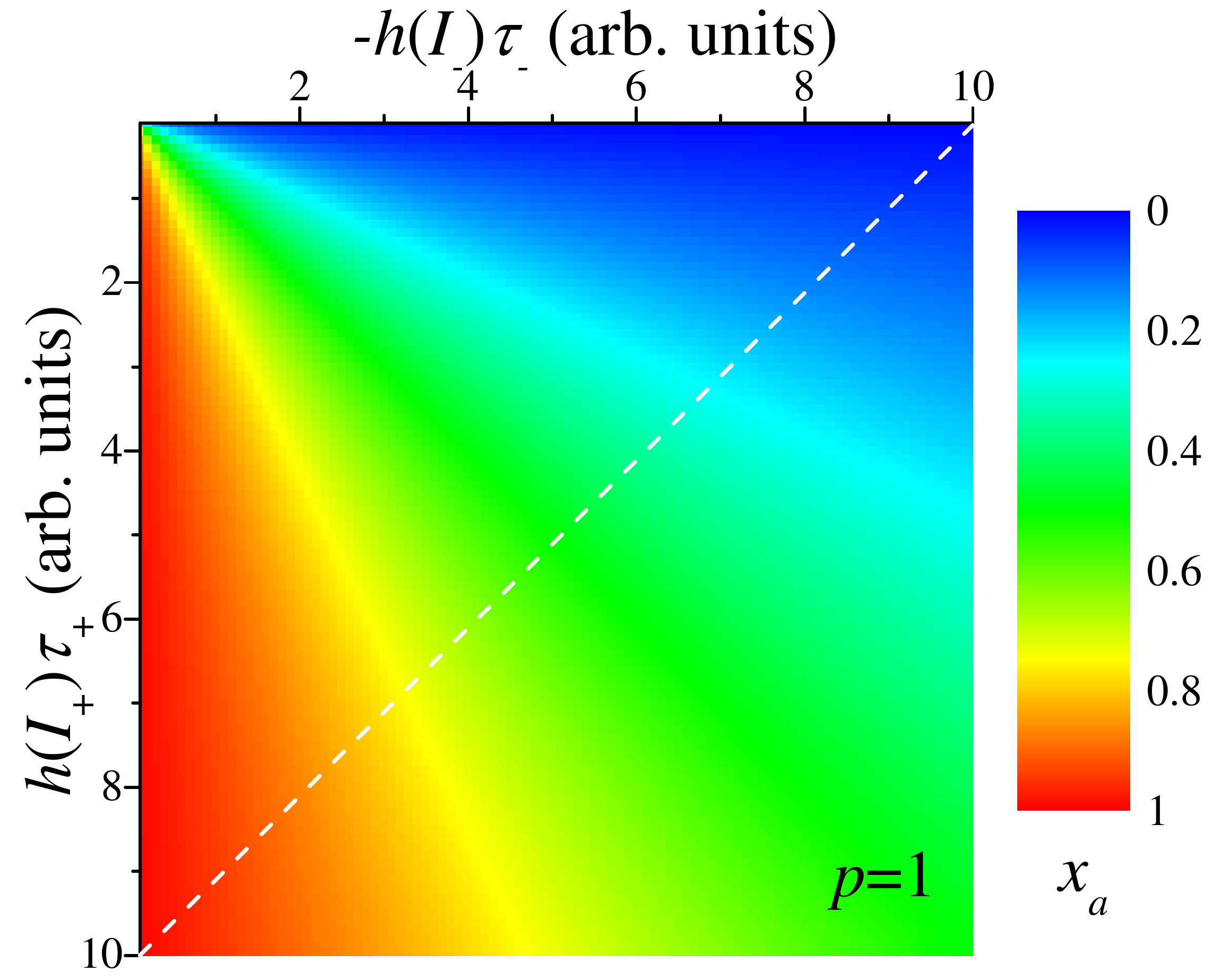} \\ (b) \includegraphics[width=70mm]{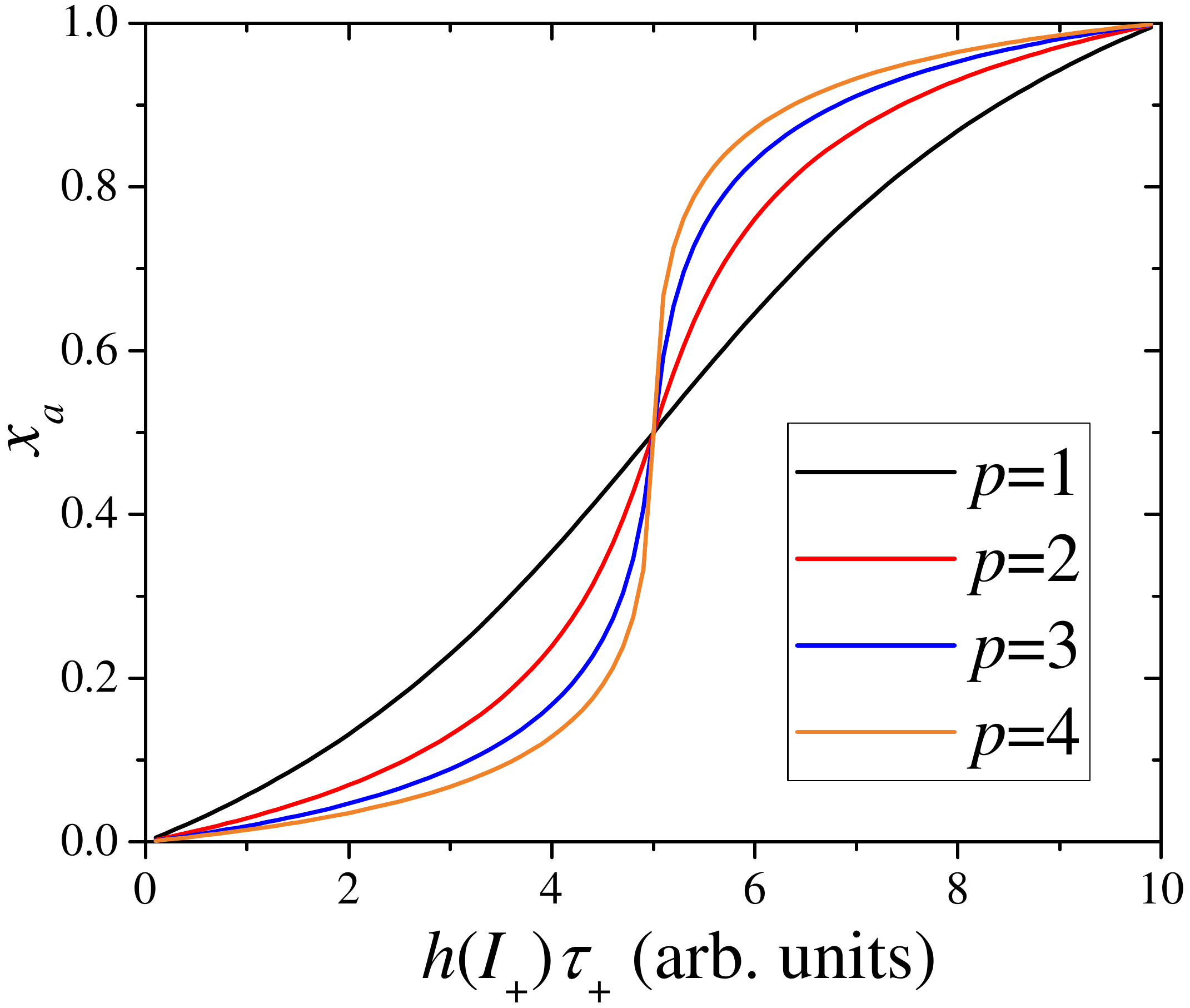}
\caption{Location of dynamical attractors in Biolek window function memristors as a function of pulse sequence parameters.
(a)~An image plot of $x_a$ for the case of $p=1$.
(b) $x_a$ as a function of $h(I_+)\tau_+$ along the diagonal dashed line in (a) for several values of $p$.
}
\label{fig:3}
\end{figure}

\begin{figure*}[t]
\centering (a) \includegraphics[width=70mm]{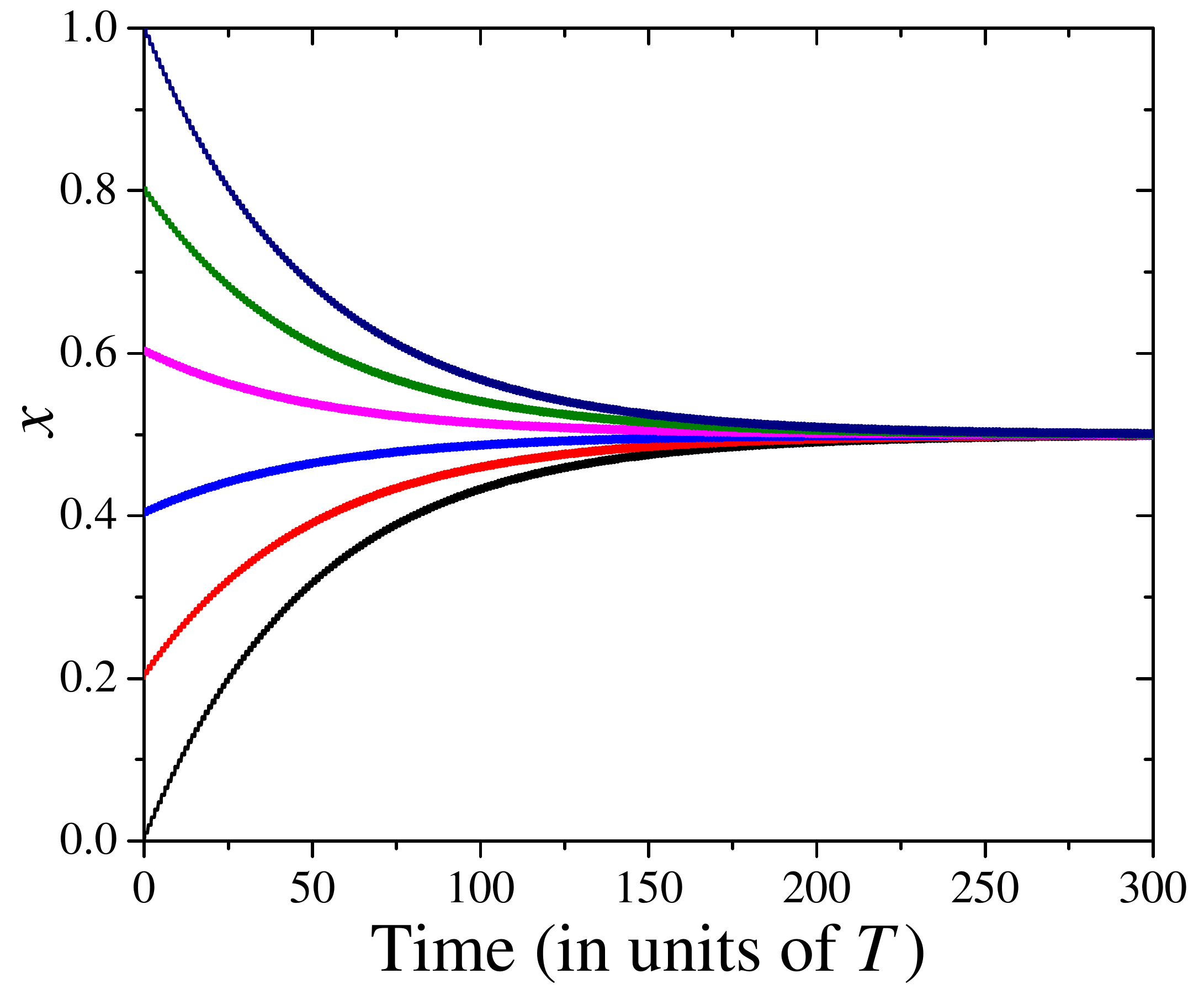} \;\;\;\;\; (b) \includegraphics[width=70mm]{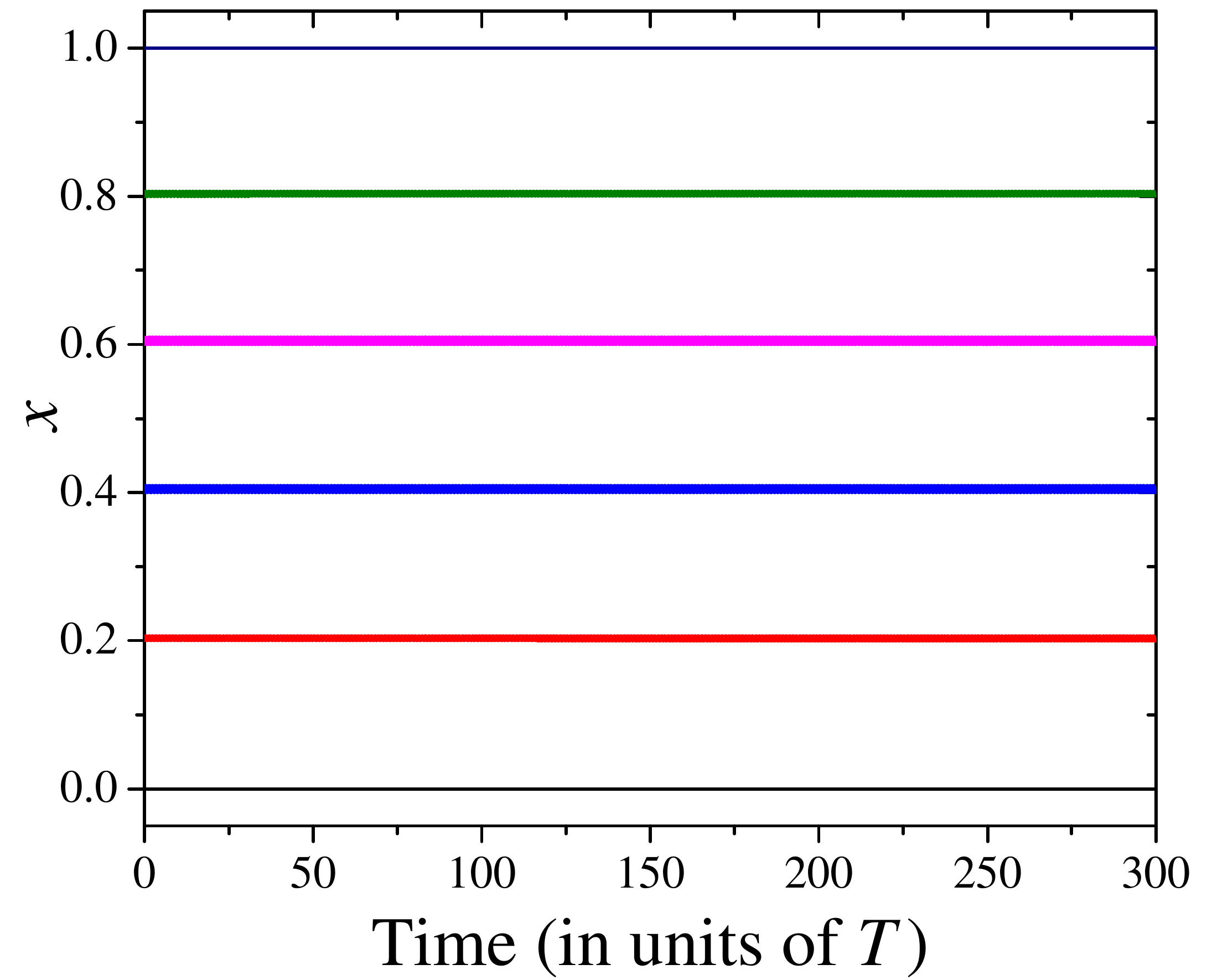}
\caption{Evolution of different initial states in the Biolek (a) and Joglekar (b) window function memristors in the case of symmetric pulses such that $h(I_+)\tau_+=-h(I_-)\tau_-$.
The fixed-point attractor dynamics in the Biolek window function memristor
contrasts with the neutral equilibrium dynamics in the Joglekar  window function memristor.
 These simulations were performed using $h(I)=\gamma I$ in Eqs.~(\ref{eq:bio_f}) and (\ref{eq:jogl_f}) written as
 $\dot{x}=\gamma I g_B(x,I)$ and $\dot{x}=\gamma I g_J(x)$, respectively,
$p=1$, $I_+=-I_-$, $\tau_+=\tau_-=0.2T$, $\gamma I_+ \tau_+ =0.01$.
}
\label{fig:4}
\end{figure*}

\textbf{Theorem 1:} There is always a single stable fixed point in the dynamics of single memristors described by Eq.~(\ref{eq:bio_f}) when these are driven by alternating polarity current pulses ($h(I_+)>0$ and $h(I_-)<0$).

\textit{Proof:} Substituting Eq.~(\ref{eq:bio_f}) into Eq.~(\ref{eq:3}) yields
\begin{equation}\label{eq:bio_pr}
  h(I_+)g(x,I_+)\tau_+=-h(I_-)g(x,I_-)\tau_-.
\end{equation}
Clearly, there exists a single solution of Eq.~(\ref{eq:bio_pr}), since the curves representing the left-hand side and right-hand side of Eq.~(\ref{eq:bio_pr}) (as functions of $x$) must intersect at some point $x_a$.
To verify that
 $x_a$ is a stable fixed point, we substitute Eq.~(\ref{eq:bio_f}) into (\ref{eq:5}) and divide it by Eq.~(\ref{eq:bio_pr}).
This leads to
\begin{equation}\label{eq:bio_pr1}
  \frac{1}{g(x,I_+)}\frac{\partial g(x,I_+)}{\partial x}< \frac{1}{g(x,I_-)}\frac{\partial g(x,I_-)}{\partial x}.
\end{equation}
As $g(x,I_+)$ is a monotonically decreasing function of $x$ while $g(x,I_-)$ is a monotonically increasing function of $x$, the left-hand side of inequality (\ref{eq:bio_pr1}) is negative and its right-hand side is positive.
Therefore, the inequality (\ref{eq:5}) is satisfied at $x_a$.

\subsection{Biolek window function memristors} \label{sec:3b}

Having established Theorem 1, consider Biolek window function memristors.
As $g_B(x,I)$, given by Eq.~(\ref{eq:bio_pr1}), satisfies the conditions listed below Eq.~(\ref{eq:bio_f}), we immediately conclude that there always exists a single fixed point in their averaged dynamics (the only required conditions are $h(I_+)>0$ and $h(I_-)<0$).
To find the point location we use Eq.~(\ref{eq:bio_pr}), which can be written as
\begin{equation}\label{eq:bio_attract}
  h(I_+)\tau_+\left( 1-x^{2p}\right)+h(I_-)\tau_-\left( 1-(x-1)^{2p}\right)=0.
\end{equation}
In the simplest case of $p=1$, the suitable solution of Eq.~(\ref{eq:bio_attract}) is
\begin{equation}\label{eq:xa}
x_a=\frac{ 1-\sqrt{\alpha^2+\alpha+1}}{\alpha +1},
\end{equation}
where $\alpha=h(I_+)\tau_+/\left( h(I_-)\tau_-\right) <0$.
For $p>1$, a numerical solution of Eq.~(\ref{eq:bio_attract}) can always be found.
Fig.~\ref{fig:3}(a) shows the location of $x_a$ corresponding to Eq.~(\ref{eq:xa}) as a function of the parameters of the pulse sequence.
This plot exhibits a clear symmetry  with respect to the parameters of the positive and negative current pulses.
The location of stable fixed point for larger values of $p$ is exhibited in  Fig.~\ref{fig:3}(b).
To generate Fig.~\ref{fig:3}(b), Eq.~(\ref{eq:bio_attract}) was solved numerically assuming $h(I_-)\tau_-=h(I_+)\tau_+-10$,
which corresponds to the dashed diagonal line in Fig.~\ref{fig:3}(a).
Fig.~\ref{fig:3}(b) demonstrates an increase in the steepness of the $x_a$ curve with $p$.
An important observation is that $x_a$ changes continuously from 0 to 1 in Fig.~\ref{fig:3}(b).
Therefore, a Biolek window function memristor can be reliably set to any desired state by an appropriate choice of the pulse sequence parameters.

It is interesting that Eq.~(\ref{eq:xa}), defining the location of attracting point $x_a$, is  `universal': all the details about the `activation' function $h(I)$ are hidden in the constant $\alpha$.
Therefore, the same $x_a$ can be realised with  different types of memristors.
For instance, one can take $h(I)=\gamma I$  (an instantaneous linear drift model) or use
 $h(I)=\textnormal{sign}\left(I\right) \gamma\left( |I|-I_{t}\right)$ if $|I|>I_t$ and $h|I|=0$ if $|I|<I_t$ (threshold-type model), etc.
Here,
 $\gamma$ is a rate constant, and $I_t$ is the current threshold.
Figure \ref{fig:4}(a) shows an example of attracting dynamics in Biolek window function memristors.
The details of the simulation are given in the caption.

It should be mentioned that the memristor potential function given by Eq.~(\ref{eq:6}) can shed some additional light on the stable fixed points.
Here we just note that in the case of Biolek window function memristors, the potential function can be written as
\begin{eqnarray}\label{eq:12}
U_B(x)&=&-h(I_+)\tau_+\left(x-\frac{x^{2p+1}}{2p+1} \right)- \\
&&\;\;\;\;\;h(I_-)\tau_-\left(x-\frac{(x-1)^{2p+1}}{2p+1} \right).
\nonumber
\end{eqnarray}

\subsection{Theorem 2} \label{sec:3c}

Next we formulate and prove a theorem concerning the absence of stable fixed points in the dynamics of memristors described by Eq.~(\ref{eq2}) of the form
\begin{equation}
  \dot{x}=h(I)g(x), \label{eq:jogl_f}
\end{equation}
where \textit{i}) $g(x)$ is a continuous function; \textit{ii}) $h(I)\geq 0$ for $I>0$, and $h(I)\leq 0$ for $I<0$; and \textit{iii}) $g(x)>0$ for $0<x<1$.

\textbf{Theorem 2:} There are no stable fixed points in the dynamics of single memristors described by Eq.~(\ref{eq:jogl_f}) when they are driven by alternating polarity current pulses ($h(I_+)>0$ and $h(I_-)<0$).
Moreover, for special cases of pulse sequences, neutral equilibrium points are possible.

\textit{Proof}: Substituting Eq.~(\ref{eq:jogl_f}) into Eq.~(\ref{eq:3}) yields
\begin{equation}\label{eq:jogl_pr}
  \left( h(I_+)\tau_+ + h(I_-)\tau_-\right) g(x)=0.
\end{equation}
Assuming that $x\neq 0$ or 1, Eq.~(\ref{eq:jogl_pr}) is satisfied only when the bracket in its left-hand side is zero.
Note that
the inequality (\ref{eq:5}) can be rewritten as
\begin{equation}\label{eq:jogl_pr1}
  \left( h(I_+)\tau_+ + h(I_-)\tau_-\right) \frac{\partial g(x)}{\partial x}<0,
\end{equation}
which, clearly, can not be satisfied simultaneously with Eq.~(\ref{eq:jogl_pr}) if we assume a zero value for the bracket.
One can recognise that the zero
bracket in Eq.~(\ref{eq:jogl_pr1}) corresponds to the condition for a neutral equilibrium.

\subsection{Joglekar window function memristors} \label{sec:3d}

In the case of  Joglekar window function memristors, Eq.~(\ref{eq:3}) takes the form
\begin{equation}
\left[h(I_+)\tau_++h(I_-)\tau_-\right]  g_J(x)=0, \label{eq:9}
\end{equation}
where $g_J(x)$ is given by Eq.~(\ref{eq:joglekar}).
It is satisfied when $h(I_+)\tau_+=-h(I_-)\tau_-$.
This condition corresponds to the neutral equilibrium point (see Fig.~\ref{fig:4}(b)).

Moreover, the  potential function (Eq.~(\ref{eq:6})) of the Joglekar window function memristor can be presented as
\begin{equation}
U_J(x)=-\left[ h(I_+)\tau_++h(I_-)\tau_-\right] \left( x-\frac{(2x-1)^{2p+1}}{2 (2p+1)}\right) \label{eq:10}.
\end{equation}
Note that $U_J(x)=0$ at the neutral equilibrium point, and $U_J(x)$ is a monotonic function of $x$ otherwise.

\begin{figure}[t]
\centering  \includegraphics[width=70mm]{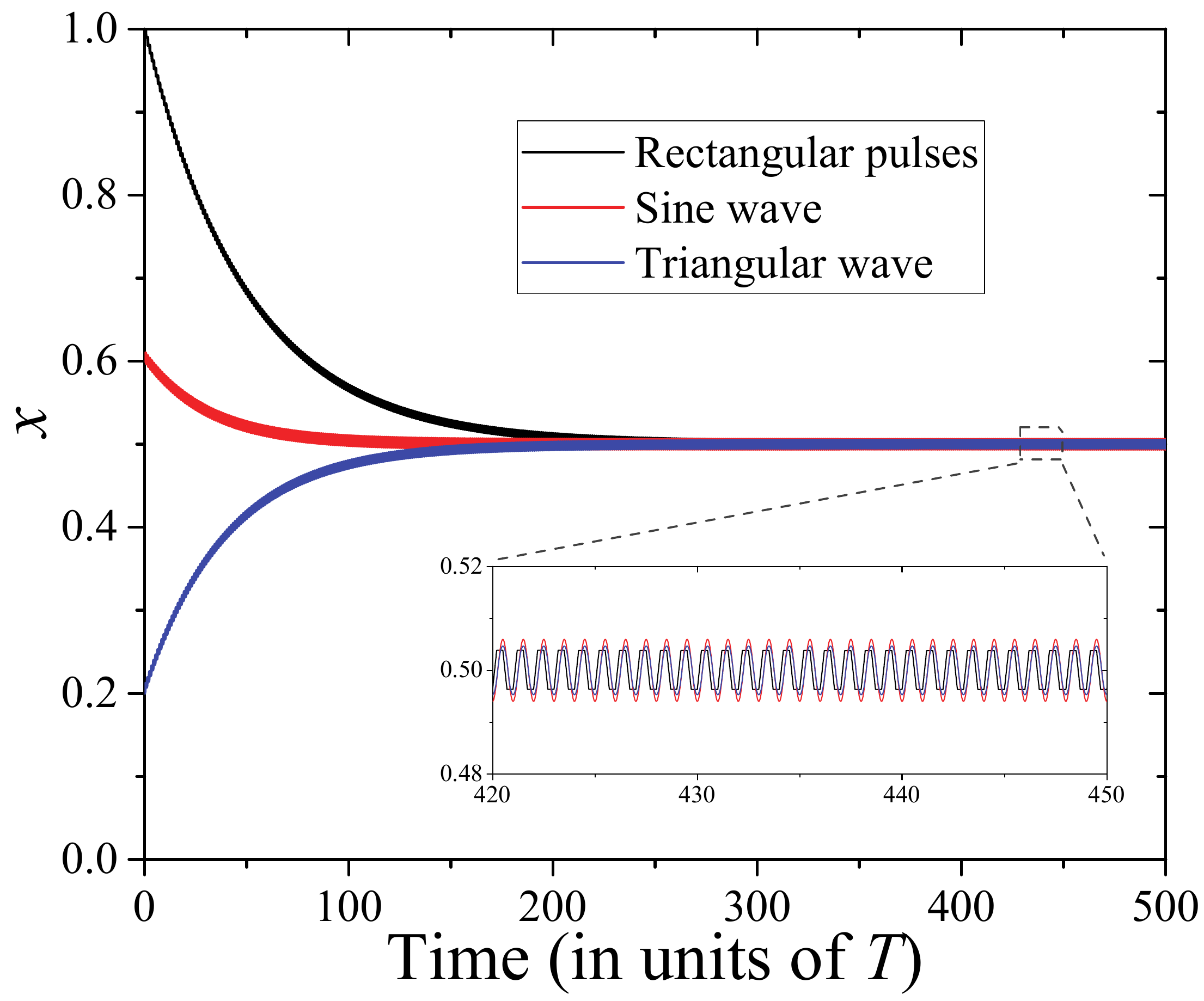}
\caption{Evolution of different initial states in the Biolek window function memristor subjected to rectangular,
sinusoidal, and triangular inputs.
These simulations were performed using the same memristor model as in Fig. \ref{fig:4} ($\dot{x}=\gamma I g_B(x,I)$) and exactly the same  rectangular pulses. The periods of sinusoidal and triangular waveforms were set equal to $T$, and their amplitudes $I_0$
are specified by $\gamma I_0 T=0.05$.}
\label{fig:5}
\end{figure}

\section{Discussion and conclusions} \label{sec:4}

Historically, window functions were introduced phenomenologically to account for the experimental observation of  boundary states of memristors (the `on' and `off' states).
To the best of our knowledge, their expressions have never been derived from fundamental physics theories.
Irrespective of this fact, several window-function based memristor models have been implemented in SPICE, and their SPICE implementations (see, for instance, Refs.~\cite{Biolek2009-1,Benderli09a,biolek2013reliable,rak2010macromodeling,yakopcic2012memristor}) are considered to be solid tools for predictive circuit simulations.
Various prospects for memristor applications have been claimed based on the results of such SPICE modelling.

In this paper we have shown that the choice of the window function is of critical importance for the predictive modelling of memristors.
In particular, it has been proven that
two broad classes of memristors (exemplified by the Biolek and Joglekar window function memristors) may demonstrate qualitatively different dynamics under the same driving conditions.
While in some practical cases this finding may be unimportant, there are situations when it can not be ignored (in particular, in circuits with memristors
subjected to periodic, quasi-periodic, and possibly random alternating polarity pulses, etc.).
Such situations are relevant to various memristor applications, including spiking neural networks and autonomous oscillating circuits, to name a few.

To show that our findings are relevant to other input signal types, we compare in Fig. \ref{fig:5} the simulated time dependence of the internal state variable in a Biolek window function memristor for the cases of rectangular, sinusoidal and triangular waveforms. In all three cases, the averaged memristor dynamics converges to the same stable fixed point. It is interesting that at the same time, the non-averaged dynamics of $x$ has the form of stable limit cycles (stable oscillations about the stable fixed point). Additionally, we note that in the case of non-linear models (when, say, $f(x,I)\sim I^2$) the attracting point locations may not be the same for different shapes of input signals.

It should be emphasised that our results are not limited to the Biolek and Joglekar window function memristors.
While the window functions described in Refs.~\cite{Biolek2009-1, Kvatinsky13a, Zha16a} satisfy the conditions of Theorem 1, the window functions introduced in Refs.~\cite{strukov08a,Benderli09a,joglekar09a,Prodromakis11a,ANUSUDHA2018130,yu2013memristor,takahashi2015spice,abdel2015memristor,georgiou16a} belong to the class covered by Theorem 2.
Moreover, it should be stressed that the experimental identification of stable fixed points could be employed to refine the models used to describe a specific type/realization of memristor. For instance, the presence or the absence of a stable fixed point in the dynamics of a physical memristor can help to select a proper window function to simulate this particular memristor.
We emphasise that our findings are also valid for voltage-controlled memristors driven by voltage pulses, with the appropriate replacement of current by voltage, and can be easily extended to other types of memory circuit elements, such as memcapacitors and meminductors~\cite{diventra09a}.

\bibliographystyle{IEEEtran}
\bibliography{memcapacitor}

\end{document}